\documentstyle[12pt]{article}
\begin{document}
\def\beq{\begin{equation}}
\def\eeq{\end{equation}}
\def\bea{\begin{eqnarray}}
\def\eea{\end{eqnarray}}
\def\ve{\vert}
\def\vel{\left|}
\def\ver{\right|}
\def\nnb{\nonumber}
\def\ga{\left(}
\def\dr{\right)}
\def\aga{\left\{}
\def\adr{\right\}}
\def\rar{\rightarrow}
\def\nnb{\nonumber}
\def\la{\langle}
\def\lla{\left<}
\def\ra{\rangle}
\def\rra{\right>}
\def\ba{\begin{array}}
\def\ea{\end{array}}
\def\tep{$B \rar K \ell^+ \ell^-$}
\def\tepm{$B \rar K \mu^+ \mu^-$}
\def\tept{$B \rar K \tau^+ \tau^-$}
\def\ds{\displaystyle}

\title{ {\Large {\bf 
The $g_{K_0^*K\pi}$ coupling constant in QCD } } }

\author{\vspace{1cm}\\
{\small T. M. Aliev \thanks
{e-mail: taliev@rorqual.cc.metu.edu.tr}\,\,,
M. Savc{\i} \thanks
{e-mail: savci@rorqual.cc.metu.edu.tr}} \\
{\small Physics Department, Middle East Technical University} \\
{\small 06531 Ankara, Turkey}
\vspace{5mm}\\
{\small F. Ya\c{s}uk}\\
{\small Physics Department, Erciyes University} \\
{\small Kayseri, Turkey} }

\date{}

\begin{titlepage}
\maketitle
\thispagestyle{empty}

\begin{abstract}
\baselineskip  0.7cm
The strong coupling constant $g_{K_0^*K\pi}$  of the scalar $K_0^*$ meson
decay to $K\pi$ is calculated in light cone QCD sum rule. The predicted 
value of the coupling constant $g_{K_0^*K\pi}$ is in a good agreement 
with the experimental result.
\end{abstract}

\vspace{1cm}
%PACS numbers: 13.20.He, 11.55.Hx, 12.38.Cy
\end{titlepage}

\section{Introduction}
The number leptons that are expected to be produced yearly at planned 
$B$ factories and the proposed $\tau$--charm factories \cite{R1}, is of the 
order of $10^7$, hence a detailed investigation of the decay properties of the 
$\tau$ lepton becomes an important issue. The decays of $\tau$ lepton can
serve not only as a useful tool in investigation of some aspects of the
standard model (SM) but also as a powerful experimental probe of new
physics \cite{R2}. CP violation plays one of the most promising role in
this direction. In light of this aspect, the decay of $\tau$ lepton into
hadrons has recently been investigated as probes of CP violation in the
scalar sector of physics beyond the SM \cite{R3}--\cite{R7}. 

In \cite{R7} the Cabibbo suppressed $\tau \rar K \pi \nu_\tau$ decay
to probe the CP violation with polarized $\tau$'s was studied. This decay
mode is dominated by the contributions of the two lowest vector $K^*$ and
scalar $K_0^*$ resonances, and the mode is expected to have larger scalar
contribution. The matrix element for $\tau \rar (K\pi)^- \nu_\tau$ in
the general form 
\bea
{\cal M} = \frac{G}{\sqrt{2}} \left[ \bar u (k) \gamma_\mu (1 - \gamma_5)
u(p) J_\mu + \bar u(p) (1 + \gamma_5) u(k) J_s \right]~,
\eea
where $p$ and $k$ are the $\tau$ lepton and the $\tau$ neutrino four
momenta, respectively. The vector and scalar hadronic matrix elements can be
parametrized as (see \cite{R7})
\bea
J_\mu &=& sin \theta_c \left< K \pi \vel \bar s \gamma_\mu u \ver 0 \right> \nnb \\ 
      &=& \sqrt{2} sin \theta_c \left[ F_K (q^2) 
\left( g_{\mu\nu} - \frac{q_\mu q_\nu}{q^2}  \right) \ga q_1 - q_2 \dr^\nu 
+ \frac{m_{K_0^*}^2}{q^2} C_K F_s (q^2) q_\mu \right]~,\nnb \\
J_s &=& \sqrt{2} sin \theta_c \left[ \frac{m_{K_0^*}^2}{m_s-m_u}\right] C_K
F_s (q^2)~,
\eea
where $\theta_c$ is the Cabibbo angle, $sin \theta_c = 0.23$, $q_1$ and 
$q_2$ are the four--momenta of $\pi$ and $K$ respectively, $m_s$ and $m_u$
are the $s$ and $u$ current quark masses and $q=q_1+q_2$ is the four--momentum of
the $K\pi$ system. The coupling strength $C_K$ denoting the scalar
contributions is determined as (see \cite{R7})
\bea
C_K = \frac{f_{K_0^*}\, g_{K_0^* K \pi}}{\sqrt{3} \, m_{K_0^*}^2}~,
\eea
where $f_{K_0^*}$ is the leptonic decay constant of scalar $K_0^*$ meson and 
$g_{K_0^* K \pi}$ is the coupling constant of the 
$K_0^* \rar K \pi$ decay. In deriving Eq. (3) we assumed
${\cal B} (K_0^* \rar K \pi) = 100\%$.
From the measured $K_0^* \rar K \pi$ decay width 
$\Gamma (K_0^* \rar K \pi) \simeq 287~MeV$, the value of $g_{K_0^* K \pi}$
is obtained to be $4.87~GeV$.   

In this work we employ light cone QCD sum rule to calculate 
$g_{K_0^* K \pi}$ coupling constant in a model independent way and compare
our results with the experimental data. 

The light cone QCD sum rule is quite different from the "classical" sum
rule which is based on the short distance operator product expansion (OPE).
This version of QCD sum rule is based on the OPE on the light cone, which
is governed by the twist of the operators rather than by their dimension and
the vacuum expectation values of local operators are replaced by the light
cone hadron wave functions, and it is quite suitable in studies of the
hard exclusive processes in QCD. 
Light cone QCD sum rule  
has been successfully applied so far in the study of many different problems
of hadron physics such as rare, radiative and semileptonic decays of $B$ meson,
$\Sigma \rar p \gamma$ decay, nucleon magnetic moment, the strong couplings 
$g_{\pi N N},~g_{\rho\omega\pi}$ and $g_{B^* B \pi}$ etc. (for an application of
this method, see for example, the recent review \cite{R8,R9} and references
therein). 

\section{QCD sum rule for the $g_{K_0^* K \pi}$ coupling constant}
The aim of this section is to calculate the coupling constant
$g_{K_0^* K \pi}$, which characterizes the $K_0^* \rar K \pi$ decay.
We start by considering the two point correlation function 
\bea
\Pi(p,q) = i \, \int d^4x \, e^{iqx} 
\lla \pi (p) \vel T \bar d(x) i \gamma_5 s(x) \, \bar s(0) u(0) \ver 0 \rra~,
\eea
which is calculated around the light cone $x^2=0$. Here 
$\bar d i \gamma_5 s$ and $\bar s u$ are the interpolating currents for
pseudoscalar $K$ and the scalar $K_0^*$ mesons, respectively. 

According to the basic idea of the QCD sum rule, we must calculate this
correlator in terms of the physical particles (hadrons) and in 
quark--gluon language, and then equate both representations.  

First let us calculate the physical part of the correlator Eq. (4). 
Saturating this correlator by $K_0^*$ and $K$ meson states, we have
\bea
\Pi^{phys.} = - g_{K_0^* K \pi} \frac{f_{K_0^*} m_{K_0^*}^2}{(m_s - m_u)} \,
\frac{f_K m_K^2}{(m_s + m_d)} \,
\frac{1}{\left( (p+q)^2 - m_{K_0^*}^2 \right)} \,
\frac{1}{(q^2 - m_K^2)}~
\eea
where $(p+q)$ and $q$ are the four momenta of the scalar $K_0^*$ and
pseudoscalar $K$ mesons, respectively.
In deriving the above equation we have used
\bea
\lla 0 \vel \bar d i \gamma_5 s \ver K \rra &=&
\frac{f_K m_K^2}{m_s + m_d}~, \nnb \\
\lla K^* \vel \bar s u \ver 0 \rra &=& i \,
\frac{f_{K_0^*} m_{K_0^*}^2}{m_s - m_u}~. \nnb \\
\eea

The strong coupling constant for the $K_0^{*-} \rar K^0 \pi^-$ decay 
is defined as follows:
\bea
\lla \pi K \ve K_0^* \rra =  - g_{K_0^* K \pi}~. \nnb
\eea

Our next task is the calculation of the theoretical part of the correlator
function (4). The full light quark propagator with both perturbative term
and contributions from vacuum fields can be written as
\bea
i {\cal S}(x) &=& \lla 0 \vel T \left\{  s(x) \bar s(0) \right\}\ver 0 \rra \nnb \\
&=& i \, \frac{\not\! x}{2 \pi^2 x^4} - \frac{\la \bar s s \ra}{12}
- \frac{x^2}{192}\, m_0^2 \la \bar s s \ra \nnb \\
&-& i \,\frac{g_s}{16 \pi^2} \int_0^1 du \Bigg\{
\frac{\not\! x}{x^2}\, \sigma_{\alpha\beta}\, G^{\alpha\beta} (ux) -
4 i u \, \frac{x_\mu}{x^2} \,G^{\mu\nu}(ux) \gamma_\nu \Bigg\}
+ \cdots~, 
\eea
where $\not\! x= x_\mu \gamma^\mu$. Note that here and in the following
formulas the strange quark mass is set zero, though in numerical analysis,
the mass of the strange quark is taken account.

Substituting Eq. (7) into correlator (4) and performing Fourier
transformation, for the theoretical part we get
\bea
\lefteqn{
\Pi^{theor.} = -f_\pi \int_0^1 du \Bigg\{ 
\varphi_\pi(u) \frac{p q}{\Delta} - 
4 \frac{p q}{\Delta^2} \Big( g_1(u) + G_2(u) \Big)
+ 2 g_2(u) \frac{1}{\Delta} }\nnb \\
&&+ \int_0^1 du \int \frac{{\cal D} \alpha_i}{\Delta_1^2} 
\Big[ \ga 2 \varphi_\perp (\alpha_i) - \varphi_\parallel (\alpha_i) +
2 \tilde \varphi_\perp (\alpha_i) - 
\tilde \varphi_\parallel (\alpha_i) \dr p q  \nnb \\
&&+ 2 u \ga \varphi_\parallel (\alpha_i) - 2 \varphi_\perp (\alpha_i) \dr
\Big] \Bigg\}~,
\eea
where 
\bea
\mu_\pi &=& \frac{m_\pi^2}{(m_u + m_d)}~, \nnb \\
\Delta  &=& - q^2 \bar u - (p+q)^2 u~, \nnb \\
\Delta_1 &=& - \left[ q+ p (\alpha_1 + u \alpha_3 ) \right]^2~, \nnb \\
{\cal D} \alpha_i &=& d \alpha_1 \, d \alpha_2 \,d \alpha_3 \,
\delta(1 - \alpha_1 - \alpha_2 - \alpha_3)~, \nnb \\
\varphi(\alpha_i) &=& \varphi (\alpha_1,\alpha_2,\alpha_3)~. \nnb \\
\eea
Here $\varphi_\pi$ is the leading twist 2 distribution amplitude, $\varphi_P$
is the two--particle distribution amplitude of twist 3; $g_1,~g_2,~
\varphi(\alpha_i)$ and $\tilde \varphi(\alpha_i)$ are the distribution
amplitude of twist 4 and 
\bea
G_2(u) = - \int_0^u g_2 (v) dv~. \nnb 
\eea
All these functions are defined as follows:
\bea
\lla \pi(p) \vel \bar d i \gamma_\mu \gamma_5 u(0) \ver 0 \rra &=&
- \,i f_\pi p_\mu \int_0^1 du e^{iupx} \ga \varphi(u) + x^2 g_1(u) \dr \nnb \\
&+& f_\pi \ga x_\mu - \frac{x^2 p_\mu}{p x} \dr 
\int_0^1 du e^{iupx} g_2(u) ~,\nnb \\ \nnb \\
\lla \pi(p) \vel \bar d i \gamma_5 u(0) \ver 0 \rra &=& 
\frac{f_\pi m_\pi^2}{(m_u+m_d)} \int_0^1 du e^{iupx} \varphi_P(u) ~,\nnb \\
\nnb \\ 
\lla \pi(p) \vel \bar d \gamma_\mu \gamma_5 g_s G_{\alpha\beta} (ux) u(0)
\ver 0 \rra &=& 
f_\pi \Bigg[ p_\beta \ga g_{\alpha\mu} - 
\frac{x_\alpha p_\mu}{px} \dr
- p_\alpha \ga g_{\beta\mu} - \frac{x_\beta p_\mu}{px} \dr \Bigg] \nnb \\
&\times& \int {\cal D} \alpha_i \varphi_\perp(\alpha_i) 
e^{i px (\alpha_1 + u \alpha_3)}~, \nnb\\ \nnb \\
\lla \pi(p) \vel \bar d \gamma_\mu  g_s \tilde G_{\alpha\beta} (ux) u(0)
\ver 0 \rra &=&  
i f_\pi \Bigg[ p_\beta \ga g_{\alpha\mu} -  
\frac{x_\alpha p_\mu}{px} \dr
- p_\alpha \ga g_{\beta\mu} - \frac{x_\beta p_\mu}{px} \dr \Bigg] \nnb \\
&\times& \int {\cal D} \alpha_i \tilde \varphi_\perp(\alpha_i) 
e^{i px (\alpha_1 + u \alpha_3)} \nnb \\
&+& i f_\pi \frac{p_\mu}{px} \ga p_\alpha x_\beta - p_\beta x_\alpha \dr 
\int {\cal D} \alpha_i \tilde \varphi_\parallel(\alpha_i) 
e^{i px (\alpha_1 + u \alpha_3)}~,
\eea
where the operator $\tilde G_{\alpha\beta}$ is the dual of
$G_{\alpha\beta}$, i.e.,
\bea
\tilde G_{\alpha\beta} = \frac{1}{2} \epsilon_{\alpha\beta\delta\rho}
G^{\delta\rho}~.\nnb
\eea
Due to the choice of the gauge $x_\mu A^\mu (x) = 0$, the path ordered gauge
factor 
\bea
{\cal P}\, exp \ga i g_s \int_0^1 du x^\mu A_\mu(ux) \dr~, \nnb
\eea
has been omitted. Note that the radiative corrections to the leading  
twist wave functions are neglected, since their contribution is small 
(about 6--7\%, see \cite{R10}).

Performing double Borel transformation with the variables $(p+q)^2$ and
$q^2$ in Eqs. (5) and (8), we get the following sum rule for the
$g_{K_0^* K \pi}$ coupling constant.
\bea
\lefteqn{
g_{K_0^* K \pi} f_{K_0^*} f_K = \frac{1}{\mu_{K_0^*}} \,
\frac{1}{\mu_K} \, 
e^{( m_{K_0^*}^2 / M_1^2 ) + ( m_K^2 / M_2^2 )} f_\pi M^4}\nnb \\
&&\times \Bigg\{- \frac{1}{2} \, \varphi_\pi^\prime(u_0)
f_1(s_0/M^2)
+ 2 \,\frac{g_1^\prime (u_0)}{M^2} 
f_0(s_0/M^2)\nnb \\
&&+ \frac{1}{M^2} \Bigg(  \int_0^{u_0} d \alpha_1
\frac{F(\alpha_1,1-u_0,u_0-\alpha_1)}{2 ( u_0 - \alpha_1)} \nnb \\
&&-\int_0^1 d \alpha_3
\frac{F(u_0,1-u_0-\alpha_3,\alpha_3)}{2 \alpha_3} \nnb \\
&&+  \int_0^{u_0} d \alpha_1 
\frac{\varphi_\parallel(\alpha_1,1-u_0,u_0-\alpha_1) - 
2 \varphi_\perp (\alpha_1,1-u_0,u_0-\alpha_1)}
{( u_0 - \alpha_1)} \nnb \\
&&-  \int_0^{u_0} d \alpha_1 \int_{u_0 - \alpha_1}^{1-\alpha_1}
\frac{1}{\alpha_3}\, \Big[ 
\varphi_\parallel(\alpha_1,1-\alpha_1 - \alpha_3,\alpha_3) 
- 2 \varphi_\perp (\alpha_1,1-\alpha_1 - \alpha_3,\alpha_3)
\Big] \Bigg) \Bigg\}~.
\eea
where 
\bea 
\mu_{K_0^*} &=& \frac{m_{K_0^*}^2}{m_s-m_u}~, \nnb \\
\mu_K &=&\frac{m_K^2}{m_s+m_d}~, \nnb \\ 
u_0 &=& \frac{M_2^2}{(M_1^2+M_2^2)}~,\nnb \\
M^2 &=& \frac{M_1^2 M_2^2}{(M_1^2+M_2^2)}~,\nnb
\eea
with $M_1^2$ and $M_2^2$ are being the Borel parameters,
\bea
\varphi_\pi^\prime (u_0) &=& \left.
\frac{d \varphi(u)}{du} \ver_{u=u_0}~~~~~\mbox{\rm and,}  \nnb \\
F(\alpha_i) &=& 2 \varphi_\perp(\alpha_i) - \varphi_\parallel (\alpha_i) +
2 \tilde \varphi_\perp(\alpha_i) - \tilde \varphi_\parallel (\alpha_i)~,
\nnb
\eea
and $s_0$ is the threshold continuum. 
Here the function
\bea
f_n(x) = 1 - e^{-x} \sum_{k=0}^n \frac{x^k}{k !}~, \nnb
\eea
is used to subtract the continuum and higher resonance contributions. 
This contribution is modeled by the dispersion integral, by invoking duality
in the region $s_1,~s_2 \ge s_0$.  
In deriving Eq. (11), we have used the double Borel transformation 
formula:
\bea
B_{1\,(p+q)^2}^{M_1^2} ~B_{2\,q^2}^{M_2^2}\, 
\frac{\Gamma(n)}{\left[ m^2 - q^2 \bar u - u (p+q)^2 \right]^n} =
\ga M^2 \dr^{2-n} e^{- m^2/M^2} \delta (u-u_0)~. \nnb 
\eea
The sum rule (11) is asymmetric with respect to the Borel
parameters $M_1^2$ and $M_2^2$ due to the significant mass difference of
$K_0^*$ and $K$. We choose $M_1^2$ and $M_2^2$ to be 
$M_1^2= 2 m_{K_0^*}^2 \beta$ and $M_2^2 = 2 m_K^2 \beta$ respectively,
where $\beta$ is a scale factor and hence in regard to this assignment we
have $M^2 = 0.44 \beta~GeV^2$, $u_0 = 0.107$.

\section{Numerical analysis}
For numerical analysis we need the explicit forms of the wave functions.
Following \cite{R11} we define the relevant wave functions as: 
\begin{equation}
\varphi_\pi(u,\mu) = 6 u(1-u)\Big[1+a_2(\mu)C^{3/2}_2(2u-1)+
 a_4(\mu)C^{3/2}_4(2u-1)\Big]
\label{expansion1}
\end{equation}
with the Gegenbauer polynomials
\bea
C_2^{3/2}(2u-1)&=&\frac{3}2[5(2u-1)^2-1]~,
\nonumber
\\
C_4^{3/2}(2u-1)&=&\frac{15}8[21(2u-1)^4-14(2u-1)^2+1]~,
\label{G2}
\eea
and the coefficients  $a_2=\frac23$, $a_4=0.43$
corresponding to the normalization point $\mu=0.5$ GeV.

\bea
\varphi_\perp (\alpha_i)&=&30\delta^2 (\alpha_1-\alpha_2)\alpha_3^2[\frac13+2
\varepsilon (1-2\alpha_3)] ~,
\nonumber
\\
\varphi_\parallel (\alpha_i)&=&120\delta^2\varepsilon (\alpha_1-\alpha_2)\alpha_1\alpha_2\alpha_3~,
\nonumber
\\
\tilde{\varphi}_\perp (\alpha_i)&=&30\delta^2\alpha_3^2(1-\alpha_3)[\frac13+2
\varepsilon (1-2\alpha_3)] ~,
\nonumber
\\
\tilde{\varphi}_\parallel (\alpha_i)&=&-120\delta^2\alpha_1\alpha_2\alpha_3[\frac13+
\varepsilon (1-3\alpha_3)] ~.
\label{tw4gluon}
\eea
\begin{eqnarray}
g_1(u)&=&\frac{5}2\delta^2\bar{u}^2u^2+\frac{1}{2}\varepsilon\delta^2[
\bar{u}u(2+13\bar{u}u)+10u^3\ln u(2-3u+\frac65u^2)
\nonumber
\\
&&{}+10\bar{u}^3\ln \bar{u}(2-3\bar{u}+\frac65\bar{u}^2)]\,,
\nonumber
\end{eqnarray}
where $\delta=0.2~GeV^2$ at $\mu=1~GeV$~. 

The values of the main input parameteres, which appear in further numerical
analysis are as follows: $\mu_\pi(1 ~GeV) \simeq 1.65,~\mu_K(1 ~GeV) 
\simeq 1,~f_\pi = 132~MeV,~f_K = 156~MeV$ and $m_s = 155~MeV$ \cite{R12}.

The dependence of $g_{K_0^* K \pi} f_{K_0^*} f_K$ on the Borel
parameter $\beta$, for different values of the threshold $s_0$ is given in 
Fig. (1). The lower bound of the Borel parameter $\beta$ is determined by
the requirement that, the terms of higher twists in the operator expansion
must be smaller than the leading twist term (say 3 times). This leads to 
$\beta \ge 1$ for the sum rule (11). The upper limit of $\beta$ is
restricted from the condition that the continuum contribution must be less
than $30\%$ of the main one. Under this condition the upper bound is
determined to be  $\beta = 1.6$.   

From the analysis of Fig. (1), we finally get
\bea
g_{K_0^* K \pi} f_{K_0^*} f_K = 0.022 \pm 0.004~,
\eea
in which we have included the uncertainties due to the continuum threshold,
Borel parameter, radiative corrections to the leading twist wave function,
neglection of the four particle components of the wave functions , 
{\it etc.} In determination of $g_{K_0^* K \pi}$ the value of the leptonic
decay constant $f_{K_0^*}$ is needed. However, it should be noted that, this
decay constant has not been measured yet, but it has been estimated in
framework of different approaches, such as QCD sum rule which predicts 
$f_{K_0^*} \simeq 31~MeV$ \cite{R13}; effective Lagrangian method whose
estimation is $\sim 45~MeV$ \cite{R14} and pole dominance model's result 
$\sim 50~MeV$ \cite{R15}. Using these values of the leptonic decay constant
$f_{K_0^*}$, we get from Eq. (12)  
\bea
g_{K_0^* K \pi} = \left\{ \begin{array}{l}
~~4.6 \pm 0.8~GeV~~~~~~\mbox {\rm (for $f_{K_0^*} = 31~MeV$)}~, \\ \\
~~3.1 \pm 0.6~GeV~~~~~~\mbox {\rm (for $f_{K_0^*} = 45~MeV$)}~, \\ \\
~~2.8 \pm 0.5~GeV~~~~~~\mbox {\rm (for $f_{K_0^*} = 50~MeV$)}~. 
\end{array} \right. \nnb
\eea
Comparing these predictions with the existing experimental result 
$g_{K_0^* K \pi}= 4.87~GeV$, it is observed that if we use 
$f_{K_0^*} = 31~MeV$, which was estimated from QCD sum rule, 
the predicted value of $g_{K_0^* K \pi}$ is
close to the experimental result. 

In summary, we have used light cone QCD sum rule to calculate the strong
coupling constant $g_{K_0^* K \pi}$. The prediction we have for 
$g_{K_0^* K \pi}$ is in good agreement with the experimental result. 

\newpage

\section{Figure captions}
{\bf Fig. 1} The dependence of $g_{K_0^* K \pi} f_{K_0^*} f_K$ on the Borel
parameter $\beta$, at two fixed values of the continuum threshold,
$s_0 = 2.4~GeV^2$ and $s_0 = 2.6~GeV^2$

\newpage

\begin{figure}
\vskip 1.5cm
    \includegraphics{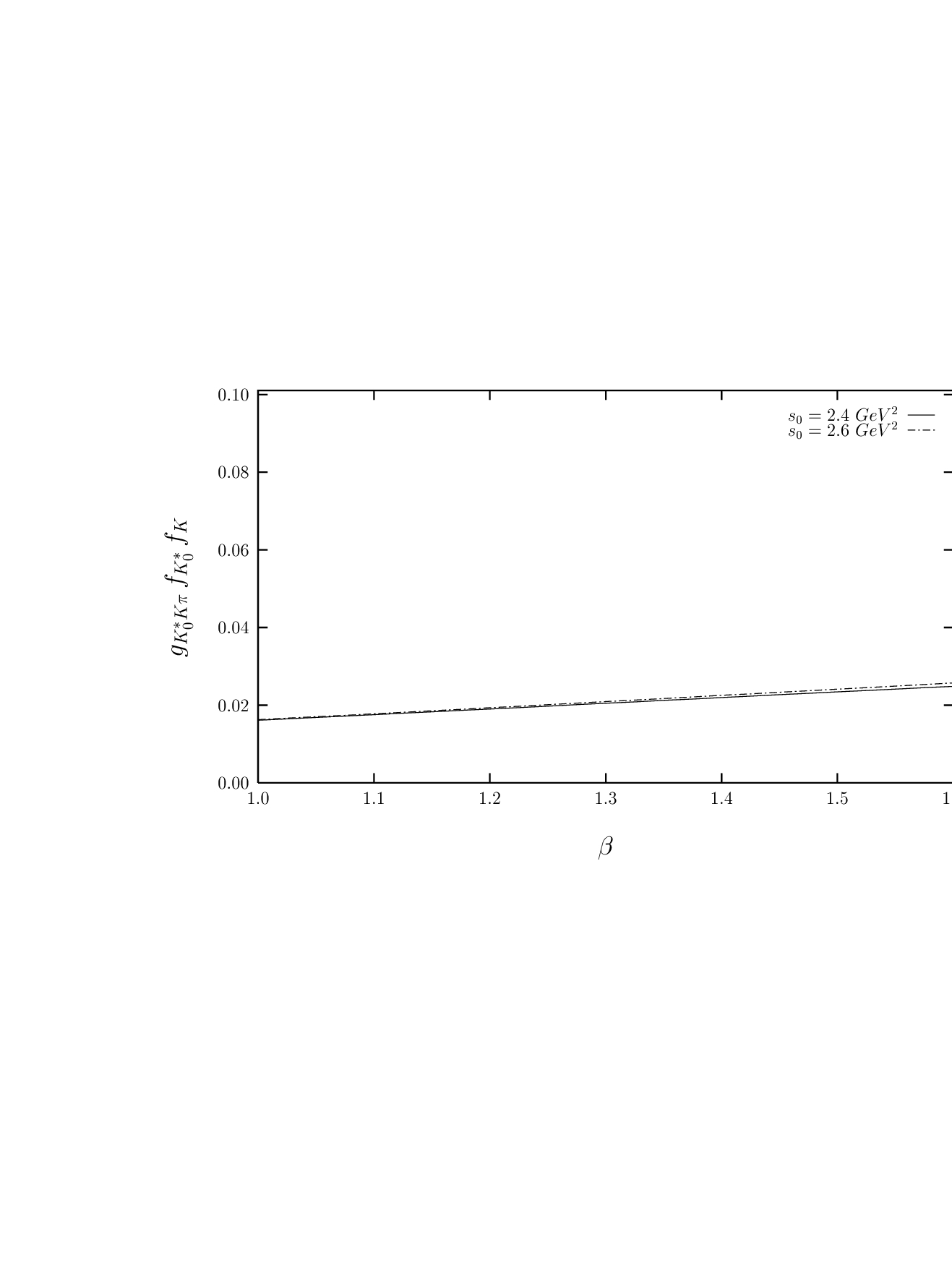}
\vskip 6.5cm
\caption{}
\vspace{10 cm}
\end{figure}

\end{document}